\documentclass[11pt,twocolumn]{IEEEtran}

\pdfoutput=1  

\IEEEoverridecommandlockouts                              
                                                          
\usepackage{graphicx,subfigure}
\usepackage{cite,url}

\usepackage[usenames,dvipsnames]{color}

\usepackage[cmex10]{amsmath}
\usepackage{amsfonts,amssymb,mathrsfs}

\usepackage[colorlinks,bookmarksopen,bookmarksnumbered,citecolor=red,urlcolor=red]{hyperref}

\def\adjsym{\star}
\def\ynom{\bar{y}^0}

\newcommand*{\qed}{\nobreak\hfill\ensuremath{\square}}

\long\def\defbox#1{\framebox[.9\hsize][c]{\parbox{.85\hsize}{%
\parindent=0pt
\baselineskip=12pt plus .1pt      
\parskip=6pt plus 1.5pt minus 1pt 
 #1}}}

\long\def\beginbox#1\endbox{\subsection*{}%
\hbox{\hspace{.05\hsize}\defbox{\medskip#1\bigskip}}%
\subsection*{}}

\def\endbox{}

\def\transpose{{\hbox{\it\tiny T}}}

\newsavebox{\junk}
\savebox{\junk}[1.6mm]{\hbox{$|\!|\!|$}}

\def\state{{\sf X}}

\newcommand{\field}[1]{\mathbb{#1}}

\def\Re{\field{R}}

\def\ind{\field{I}}

\def\Co{\field{C}}

\def\bfmath#1{{\mathchoice{\mbox{\boldmath$#1$}}%
{\mbox{\boldmath$#1$}}%
{\mbox{\boldmath$\scriptstyle#1$}}%
{\mbox{\boldmath$\scriptscriptstyle#1$}}}}

\def\bfmX{\bfmath{X}}

\def\bfmY{\bfmath{Y}}

\def\bfmhhaY{\bfmath{\hhaY}} 
\def\bfmhhaY{\hbox to 0pt{$\widehat{\bfmY}$\hss}\widehat{\phantom{\raise 1.25pt\hbox{$\bfmY$}}}}

\def\til={{\widetilde =}}

\def\clX{{\cal X}}

\def\head#1{\subsubsection*{#1}}

 \def\FRAC#1#2#3{\genfrac{}{}{}{#1}{#2}{#3}}

\def\ddt{{\mathchoice{\FRAC{1}{d}{dt}}%
{\FRAC{1}{d}{dt}}%
{\FRAC{3}{d}{dt}}%
{\FRAC{3}{d}{dt}}}}

\def\ddtp{{\mathchoice{\FRAC{1}{d^{\hbox to 2pt{\rm\tiny +\hss}}}{dt}}%
{\FRAC{1}{d^{\hbox to 2pt{\rm\tiny +\hss}}}{dt}}%
{\FRAC{3}{d^{\hbox to 2pt{\rm\tiny +\hss}}}{dt}}%
{\FRAC{3}{d^{\hbox to 2pt{\rm\tiny +\hss}}}{dt}}}}

\def\ddzeta{{\mathchoice{\FRAC{1}{d}{d\zeta}}%
{\FRAC{1}{d}{d\zeta}}%
{\FRAC{3}{d}{d\zeta}}%
{\FRAC{3}{d}{d\zeta}}}}

\def\ddt{\frac{d}{dt}}

\def\ddzeta{\frac{d}{d\zeta}}

\def\eqdef{\mathbin{:=}}

\def\Expect{{\sf E}}

\def\average#1,#2,{{1\over #2} \sum_{#1}^{#2}}

\def\eye(#1){{\bf(#1)}\quad}

\def\varble{\,\cdot\,}

\newtheorem{theorem}{Theorem}[section]

\newtheorem{proposition}[theorem]{Proposition}
\newtheorem{lemma}[theorem]{Lemma}

\def\Proposition#1{Proposition~\ref{#1}}

\def\Section#1{Sec.~\ref{#1}}

\def\eq#1/{(\ref{e:#1})}

\newcommand{\eeqn}{\end{eqnarray} }

\newcommand{\eeq}{\end{equation}}

\def\bdes{\begin{description}}
\def\edes{\end{description}}

\newcounter{rmnum}
\newenvironment{romannum}{\begin{list}{{\upshape (\roman{rmnum})}}{\usecounter{rmnum}
\setlength{\leftmargin}{14pt}
\setlength{\rightmargin}{8pt}
\setlength{\itemsep}{2pt}
\setlength{\itemindent}{-1pt}
}}{\end{list}}

\newcounter{anum}

{\end{list}}

\def\ass(#1:#2){(#1\ref{#1:#2})}

\def\ritem#1{
\item[{\sf \ass(\current_model:#1)}]
}

\newenvironment{recall-ass}[1]{%
\begin{description}
\def\current_model{#1}}{
\end{description}
}

\def\Ebox#1#2{%
\begin{center}
 \parbox{#1\hsize}{\epsfxsize=\hsize \epsfbox{#2}}
\end{center}}

\newcommand{\bd}{\begin{description}}
\newcommand{\ed}{\end{description}}
\newcommand{\bt}{\begin{theorem}}
\newcommand{\et}{\end{theorem}}
\newcommand{\ba}{\begin{array}{rcl}}
\newcommand{\ea}{\end{array}}

\def\head#1{\paragraph{#1}}

\newlength{\noteWidth}
\long\def\notes#1{\ifinner
           {\tiny #1}
           \else
           \marginpar{\parbox[t]{\noteWidth}{\raggedright\tiny #1}}
       \fi\typeout{#1}}

\def\notes#1{}  

\makeindex
 
\def\Ebox#1#2{%
\begin{center}    
\includegraphics[width=#1\hsize]{#2}
\end{center}}

\def\util{\mathchoice{\mbox{\small$\cal U$}}%
{\mbox{\small$\cal U$}}%
{\mbox{$\scriptstyle\cal U$}}%
{\mbox{$\scriptscriptstyle\cal U$}}}

\def\tilutil{\mathchoice{\mbox{\small$\cal \widetilde U$}}%
{\mbox{\small$\cal\widetilde U$}}%
{\mbox{$\scriptstyle\cal \widetilde U$}}%
{\mbox{$\scriptscriptstyle\cal \tilde U$}}}

\def\welf{\mathchoice{\mbox{\small$\cal W$}}%
{\mbox{\small$\cal W$}}%
{\mbox{$\scriptstyle\cal W$}}%
{\mbox{$\scriptscriptstyle\cal W$}}}
 
\def\cp{{\check{p}}}
\def\cL{{\check{L}}}

\def\cgenerate{{\check{\generate}}}

\def\Fig#1{Fig.~\ref{#1}}

\def\Prop#1{Prop.~\ref{#1}}     
\def\Thm#1{Thm.~\ref{#1}}

 \def\Real{\text{Re}\,}

 \def\diag{\text{diag}\,}

\def\generate{{\cal D}}

\title{\LARGE \bf
Passive Dynamics in Mean Field Control
}

\author{Ana Bu\v{s}i\'c and Sean Meyn
\thanks{This research is supported by the French National Research Agency grant ANR-12-MONU-0019,  NSF grants CPS-0931416 and  CPS-1259040,   and  US-Israel BSF Grant 2011506.}
\thanks{A.B.\ is with Inria and the Computer Science Dept. of \'Ecole Normale Sup\'erieure, Paris, France.}%
\thanks{
S.M. is with the Department of Electrical and Computer
Engg.\ at the University of Florida, Gainesville.}%
}

\begin{document}

\maketitle
\thispagestyle{empty}
\pagestyle{empty}


\begin{abstract} 

Mean-field models are a popular tool in a variety of fields.  They provide an understanding of  the
impact of interactions among a large number of particles or people or other ``self-interested agents'', and are an increasingly popular tool in distributed control.

This paper considers a particular randomized distributed control architecture introduced in our own recent work. In numerical results it was found that  the associated mean-field model had attractive properties for purposes of control.  In particular, when viewed as an input-output system, its linearization was found to be minimum phase.

In this paper we take a closer look at the control model.  The   results are summarized as follows:
\begin{romannum}
\item 
The Markov Decision Process framework of Todorov is extended to continuous time models,  in which the ``control cost'' is based on relative entropy.  This is the basis of the construction of a family of Markovian generators,  parameterized by a scalar $\zeta\in\Re$.

\item 
A decentralized control architecture is proposed in which each agent evolves as a controlled Markov process.   A central authority broadcasts a common control signal $\{\zeta_t\}$ to each agent.    The central authority chooses $\{\zeta_t\}$  based on an  aggregate scalar output of the Markovian agents.

\textit{This is the basis of the mean field model.  }

\item
Provided the control-free system (with $\zeta\equiv 0$) is a reversible Markov process, the following identity holds for the transfer function $G$ obtained from the linearization,  
\[
\Real (G(j\omega)) = \text{PSD}_Y(\omega)\ge 0 \qquad \omega\in\Re\,,
\]
where the right hand side denotes the power spectral density for the output of any one of the 
individual Markov processes  (with $\zeta\equiv 0$).  
\end{romannum}

\end{abstract}


\section{Introduction}

Mean field models are a standard  tool in physics when analyzing a large number of particles, where an individual particle has negligible impact upon the ensemble.  Similar models are the foundation of competitive equilibrium theory in economics, and mean field models are increasingly popular in   control theory  \cite{huacaimal07,borsun12,gasgauleb12,guaragwil12}.

The present work considers application to distributed control, inspired by numerical results  in our prior work \cite{meybarbusyueehr14} on automated demand response for a large collection of loads.  The goal was to obtain \textit{ancillary service} to help regulate the power grid, as in many prior works \cite{coupertemdeb12,macalhis10}.

The paper \cite{meybarbusyueehr14}  focused on a large population of ``on-off'' loads, with special attention to residential pool pumps.   The normal operation of a pool pump was modeled as a Markov decision process, which included as an exogenous input a regulation signal from a balancing authority.   This resulted in an input-output system with input equal to the regulation signal, and output equal to the number of pools in operation.  In the numerical example considered, the control system had some very attractive properties:  Its linearization was stable, and simulations of one million pools resulted in behavior very closely matched to the deterministic linear model obtained from linearization of the Markovian dynamics.  Most important was the finding that the linearization was \textit{minimum phase}.  This is a valuable property in any control system.  

In this paper we set out to see why these conclusions might be expected in greater generality.

\begin{figure}[h]
\Ebox{.9}{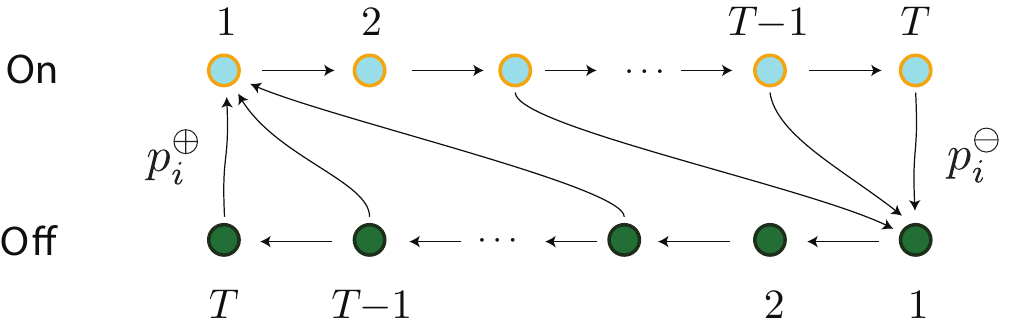} 
\vspace{-.25cm}
\caption{State transition diagram for the pool-pump model.
}
\label{fig:pppDynamics} 
\end{figure}

To explain the goals of the paper we take a high-level look at the prior work \cite{meybarbusyueehr14}.
 Shown in \Fig{fig:pppDynamics} is a state transition diagram for the discrete-time Markovian model considered in \cite{meybarbusyueehr14}.  
 The variables $p^\oplus$  and $p^\ominus$
indicate the probability of turning a pool pump on (respectively, off),  which depends upon how long the pool has been off (respectively, on).

A continuous time counterpart is described by 
a model on a continuous state space, 
\[
\state=\{  (m,\tau) :  m\in  \{  \oplus,\ominus\}  ,\  \tau\ge 0 \}  .
\]
If $X_t = (\oplus, \tau)$,  this means that the pool pump has been operating for exactly $\tau$ seconds.
If $t>\tau$, this implies that the pump  was turned on at time $t-\tau$.
The differential generator for this Markovian model is defined  for functions $f\colon\state\to\Re$ that are differentiable in their second variable.  There are functions $q^\oplus(\varble)$ and $q^\ominus(\varble)$ such that for any such $f$,
\[
\generate f\, (x) =   
\begin{cases}
q^\ominus(\tau) [ f(\ominus,0) - f(\oplus,\tau) ] + \frac{\partial}{\partial \tau} f\, (\oplus,\tau), \qquad \quad\\ 
\hfill x= (\oplus,\tau)
\\[.2cm]
q^\oplus(\tau) [ f(\oplus,0) - f(\ominus,\tau) ] + \frac{\partial}{\partial \tau} f\, (\ominus,\tau) , \\
\hfill x= (\ominus,\tau)
\end{cases}
\]
Hence $q^\oplus(\tau)$ is the \textit{jump rate} to the on-state, for a pool that has been off for $\tau$ seconds.

In this prior work the Markovian dynamics were controlled through a signal $\{\zeta_t\}$ that is broadcast to all pools. For this continuous time model, the jump rates would be modified by this signal.  With $N$ pools, on  letting  $ X^i_t  $ denote the state of the $i$th pool, the following limit is shown to  hold under mild assumptions:
\begin{equation}
\lim_{N\to\infty}
\frac{1}{N}\sum_{i=1}^N  \ind\{ X^i_t \in A \} = \mu_t(A)\,, \quad A\subset \state.  
\label{e:mfgPool}
\end{equation}  
If $\generate_{\zeta_t}$ denotes the transformed generator at time $t$, then the limit is the solution to the differential equation,
\begin{equation}
\ddt \mu_t = \mu_t \generate_{\zeta_t}
\label{e:muState}
\end{equation}
This means that for functions $f$ satisfying the conditions above,
\[
\ddt \int f(x)\mu_t(dx) =   \int  \bigl( \generate_{\zeta_t}
f\, (x)\bigr)\mu_t(dx) \, .
\]
The output is defined by a linear function of $\mu_t$: For a function $\util\colon\state\to\Re$,
\begin{equation}
y_t =   \int \util(x)\mu_t(dx)  
\label{e:muOutput}
\end{equation}
The coupled equations (\ref{e:muState},\ref{e:muOutput}) describe a nonlinear input-output model of the form considered in this paper. The input $\zeta_t$ and output $y_t$ are assumed to be real-valued.

Because \eqref{e:muState} is linear in the ``state'' $\mu_t$,  and \eqref{e:muOutput} is also linear in $\mu_t$,  it is easy to obtain a linearized model given some structure on the controlled generator.  The question addressed in this paper is, \textit{why should the linearized system have good properties for the purposes of control}?

We address this question for models in continuous time, since 
the analysis is most elegant in this setting.  While many of the results in this paper can be extended to a general state space setting,   for the remainder of the paper we restrict to a finite state space,  $\state=\{ x^1,\dots,x^d\}$.   The family of generators $\{\generate_\zeta :\zeta\in\Re\}$ is a collection of $d\times d$ matrices, which are assumed to be a smooth function of the scalar parameter $\zeta$.

The linear model is  intended to approximate the nonlinear model near an equilibrium.   To define the equilibrium we let $\pi$ denote an invariant probability measure for the Markov process with generator $\generate =\generate_0$.  
This satisfies the invariance equation,
\[
\pi \generate\, (x^j) = \sum_{i=1}^d \pi(x^i) \generate(x^i, x^j) = 0,\qquad x^j\in\state.
\]
For the nonlinear model \eqref{e:muState}, 
if $\zeta_t\equiv 0$ and if $\mu_0=\pi$, then $\mu_t=\pi$ for all $t$.  
 
The linear model evolves according
to the $d$-dimensional  linear state space equations,
\begin{equation}
\ddt \Phi_t = A \Phi_t + B \zeta_t,\qquad \gamma_t = C \Phi_t 
\label{e:LSSmfg}
\end{equation}
The $i$th component of $\Phi_t$ is intended to approximate $\mu_t (x^i) - \pi(x^i)$.
The $d\times d$  matrix $A$ is the transpose of $\generate_0=\generate$.

The $d$-dimensional row vector $C$ has components 
\begin{equation}
C_i = \tilutil(x^i) =  \util(x^i) -\ynom,
\label{e:Cdefn}
\end{equation}
where $\ynom = \sum_j \pi(x^j) \util(x^j)$.  
The output $\gamma_t =C\Phi_t$ is intended to approximate $y_t - \ynom$.
 
Letting  $\generate'_0(x^i, x^j) $ denote the derivative of $\generate_\zeta(x^i, x^j) $ with respect to $\zeta$, evaluated at $\zeta=0$,  the $d$-dimensional column vector $B$ can be expressed,
\begin{equation}
B_j = \sum_{i=1}^d \pi(x^i) \generate'_0(x^i, x^j) ,\quad 1\le j\le d.
\label{e:B}
\end{equation}

The transfer function for this model is $G(s) = C[Is - A]^{-1}B$, $s\in\Co$.  The minimum phase condition means that all zeros of $G$ lie in the strict left half plane.  In this paper we establish a stronger condition on the transfer function:
Under the assumption that the nominal Markov model is \textit{reversible}, the
 linear dynamics satisfy the \textit{positive real} condition,
\notes{why no zeros?  We may only have PR.}
\begin{equation}
\Real (G(j\omega)) \ge 0,\qquad \omega\in\Re.   
\label{e:spr}
\end{equation}

We obtain positivity by establishing the following identity,
\begin{equation}
\Real (G(j\omega)) = \text{PSD}_Y(\omega) \qquad \omega\in\Re\,,
\label{e:sprPSD}
\end{equation}
where the right hand side denotes the power spectral density for $\{ Y_t= \util(X_t) \}$ with $\bfmX$ the stationary Markov process with generator $\generate$.

The positive real condition is established only when the family of generators is constructed using the optimal control approach described in \Section{s:meanfield}.   This recalls a similar result from linear optimal control theory, where it is known that the positive real condition holds for a certain transfer function, provided the system is controlled using state feedback based on linear-quadratic optimal control \cite{kal64,safath77}.  We do not know if there is any connection between the main results of this paper, and these celebrated results from linear control theory.
   
The remainder of the paper is organized as follows:  \Section{s:MDP}
contains an extension of Todorov's optimal control framework to Markovian models in continuous time.  This is the basis of the mean-field model in \Section{s:meanfield}, and the main result that establishes the identity \eqref{e:sprPSD}.
An example  is given to show that reversibility of the nominal model is necessary in general.  Conclusions and   discussion are contained in  \Section{s:conc}.
 
\section{Construction of the controlled generator}
\label{s:MDP}

Here we describe a stochastic optimal control problem  in which the input is completely unconstrained.  The optimization criterion will include a scalar weighting term $\zeta$.  The optimal solution will define the generator $\generate_\zeta$ that is used in the mean field analysis that follows.

We consider a model in continuous time, with finite state space  $\state=\{ x^1,\dots,x^d\}$.  The optimization is based on a \textit{nominal}  Markov process on this state space.
Its generator (i.e.,  rate matrix) is defined for functions $f\colon\state\to\Re$ via, 
\begin{equation}
\begin{aligned}
\generate f\, (x)  &= \sum_{x'} \generate(x,x') f(x') \\
& = \lim_{t\downarrow 0} \frac{1}{t} \Expect[f(X_t) - f(X_0) \mid X_0=x]\,, \quad x\in\state.
\label{e:gendef}
\end{aligned}
\end{equation}
The transition semigroup is the exponential, $P^t = e^{t\generate}$, $t\ge 0$.
The Markov process is assumed to be irreducible, so that there is a unique invariant probability measure $\pi$:  Interpreted as a row vector, it satisfies   $\pi \generate =0$  and $\pi P^t=\pi$ for $t\ge 0$.

For fixed $T$ and fixed initial condition $X(0)=x$, let $p^0$ denote the probability 
distribution for the stochastic process $\{X_t: 0\le t\le T\}$ for the nominal model.

This is an unusual stochastic control problem because there is no explicit ``input''.   Any modification $p$ of $p^0$ is permitted.   A particular optimization objective will ensure that an optimal solution is Markovian.

It is assumed that a utility function $\util\colon\state\to\Re$  is given,  that represents some benefit as a function of state. The cost of deviation $p\neq p^0$ is defined by Kullback-Leibler divergence, denoted $ D(p \| p^0)$.
The $T$-stage welfare is defined as the difference,
\begin{equation}
\label{e:Twelfare}
\welf_T(p) =    \zeta \Expect_p\Bigl[\int_0^T \util(X_t)\, dt\Bigr] -D(p \| p^0)
\end{equation}
where in the expectation $\{X_t: 0\le t\le T\}$ is distributed according to $p$. 
The maximizer exists, and is denoted $p^*$ (or $p^*_T$ when the time horizon is emphasized).

The parameter $\zeta$ is a real scalar.
For notational simplicity, until \Section{s:meanfield}
 we take $\zeta=1$.

Before proceeding with the formula for $p^*$, 
it is helpful to recall the definition of divergence in this sample-path setting. 
We let $\clX_T$ denote the sigma algebra generated by the stochastic process $\{X_t: t\le T\}$.
A log-likelihood ratio is interpreted as an $\clX_T$-measurable random variable:  If  $p$  admits a log likelihood ratio $L$, this means that for any $\clX_T$-measurable random variable $F$ 
we can write, 
\begin{equation}
\Expect_p[F] = \Expect[e^L F]
\label{e:likely}
\end{equation}
where the expectation on the left is under $p$, and the expectation on the right is under  $p^0$ (the subscript is not used for the nominal model).  The K-L divergence is then defined to be,
\[
D(p \| p^0) = \Expect_p[L]
\]
If $L$ does not exist, then $D(p \| p^0) =\infty$.

\begin{proposition}
\label{t:checkp}
Suppose that  the nominal model
is irreducible, and that $X(0)=x$ is specified. 
Then $p^*_T$ is unique, and is given by the twisted  (or `tilted') distribution that is uniquely defined by the log likelihood ratio,
\begin{equation}
L^* =  {- \Lambda^*_T}  + \int_0^T  \util(X_t) \, dt 
\label{e:Lstar}
\end{equation}
The optimal welfare $\welf_T( p^*_T)   $
coincides with the constant $\Lambda^*_T$ appearing in \eqref{e:Lstar}, which is equal to the cumulative log-moment generating function,
 \begin{equation}
 \Lambda^*_T 
 = \log \Bigl(\Expect \Bigl[ \exp\Bigl(\int_0^T  \util(X_t) \, dt\Bigr) \Bigr]\Bigr)
\label{e:LambdaNorm1}
\end{equation}
where the expectation is w.r.t.\ the nominal model.
\end{proposition}

\head{Proof of \Proposition{t:checkp}}

Optimality of $p^*$ (with log likelihood ratio \eqref{e:Lstar})
is a consequence of Kullback's inequality (see eqn (4.5) of \cite{kul54}).
See also  Theorem 3.1.2  of  \cite{demzei98a}  for a version of this result on a finite probability space.
The papers \cite{pramenrun96,huaunnmeyveesur11} contain more background and other applications of this result.  

\notes{we can leave this out:
}

An explicit value for the optimal welfare follows:  We have,
\[ 
D( p^* \| p) =    \Expect_{p^*}[L^*] = -\Lambda^*_T  +\Expect_{p^*}\Bigl[\int_0^T \util(X_t)\, dt\Bigr]
\]
and consequently,
\[
\max_{p} \welf_T(p) =\welf_T( p^*) = \Lambda^*_T
\] 

The formula \eqref{e:LambdaNorm1} follows from the fact that  $ p^*=e^{L^*} p^0$ defines a probability distribution:
\begin{equation}
1= \Expect [e^{L^*}]=
e^{-\Lambda^*_T}\Expect \Bigl[ \exp\Bigl(\int_0^T  \util(X_t) \, dt\Bigr) \Bigr]
\label{e:LambdaNorm}
\end{equation}
\qed

While the optimal probability measure $ p^*$ is Markovian, it is not time-homogeneous.

We now consider an infinite horizon optimization problem:  Find a Markov process 
for which the associated family of distributions $\{\cp_T\}$,  
with initial condition $X(0)=x$,  attains the limit,
\begin{equation}
\lim_{T\to\infty} \frac{1}{T} \welf_T( \cp_T)  
=
\welf_\infty^*
\eqdef 
\lim_{T\to\infty} \frac{1}{T} \welf_T( p^*_T)   
\label{e:Cinfty1}
\end{equation}
This has a solution defined by a time-homogeneous Markov process, whose generator is denoted $\cgenerate$.   Its construction is based on the solution to an eigenvector problem: Let $I_{\util}$ denote the diagonal matrix $I_{\util} = \diag(\util(x^1),\dots, \util(x^d))$,
and let $v\colon\state\to(0,\infty)$ denote a non-trivial solution to the eigenvector problem,
\begin{equation}
[\generate + I_{\util}] v = \Lambda v
\label{e:evector}
\end{equation} 
where $\Lambda$ is the eigenvalue of $\generate + I_{\util}$ with maximal real-part (the Perron-Froebenius eigenvalue \cite{konmey05a}).

\begin{proposition}
\label{t:checkgenerate}
The following hold under the assumptions of 
\Proposition{t:checkp}:
\begin{romannum}

\item 
$\welf_\infty^*=
\Lambda $; the eigenvalue appearing in \eqref{e:evector}.

\item The generator for the Markov process that attains the optimal average welfare $\welf_\infty^*$ is obtained by  normalizing $\generate + I_{\util}$,  and applying a similarity transformation using $I_v$:
\begin{equation}
\cgenerate =   \ind_{v}^{-1} [\generate + I_{\util} - \Lambda  I]\ind_{v}
\label{e:cgenerate}
\end{equation}

\item
For each $T$,  the welfare for the distribution $\cp_T$ is given by,
\[
\welf_T(\cp_T)   =  T\welf_\infty^* - \Expect\Bigl[\log\Bigl( \frac{v(X_T)}{v(X_0)} \Bigr)  \Bigr]   
\]
\qed
 \end{romannum}
\end{proposition}

The eigenvector equation \eqref{e:evector} implies that $\sum_{x'} \cgenerate (x,x') =0$ for all $x\in\state$, as required for a Markovian generator. 


   \begin{figure*}
\Ebox{}{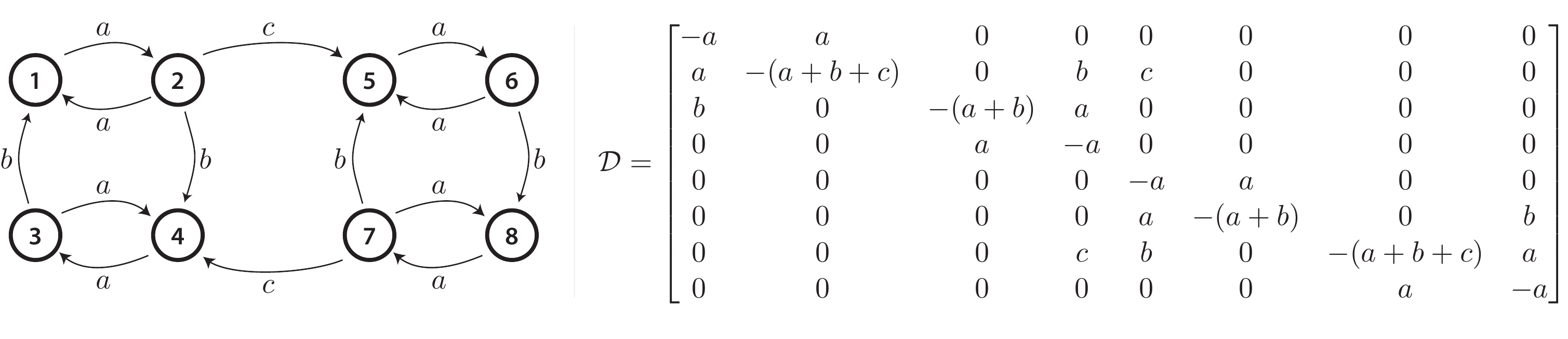}
\vspace{-.35cm}
\caption{State transition diagram and generator for a Markov process that is not reversible.  The transfer function for the linearized mean field model is neither positive real nor minimum phase.} 
\label{fig:AB} 
\end{figure*} 

\notes{"As in Todorov  \cite{tod07}".  What did he actually show!? }

\head{Proof of \Proposition{t:checkgenerate}}

Part (i) is essentially known:
From \eqref{e:LambdaNorm} it follows that $\welf_\infty^*$ is the multiplicative-ergodic limit,
\begin{equation} 
\welf_\infty^* =
\lim_{T\to\infty} \frac{1}{T} \log\Bigl( \Expect \Bigl[ \exp\Bigl(\int_0^T  \util(X_t) \, dt\Bigr) \Bigr]  \Bigr)
\label{e:CinftyLambda}
\end{equation}     
The right hand side is denoted  $ \Lambda(\util)$ in  \cite{konmey05a}, where it is shown in far greater generality that  the multiplicative-ergodic limit  $ \Lambda(\util)$ coincides with the eigenvector $\Lambda$.

The proof that $\Lambda= \Lambda(\util)$ and the proof of the remaining claims of the theorem will follow from a representation of the Markov process with generator $\cgenerate$.

Let $\cp$ denote the probability measure on sample paths, with given initial condition $X(0)=x$.  For finite $T$,  if $F$ is an $\clX_T$-measurable functional then,
for the Markovian model with generator $\cgenerate$ we have (exactly as in \eqref{e:likely}),
\[
\Expect_{\cp}[F] = \Expect\bigl[ e^{\cL_T}F\bigr]
\]
where the expectation is with respect to $p^0$, and   
\[
\cL_T = \log\Bigl( \frac{v(X_T)}{v(X_0)} \Bigr) +  \int_0^T[  \util(X_t) -\Lambda ]\, dt 
\]
Using the fact that $\cp_T$ is a probability distribution gives,
\[
1= \Expect [e^{\cL_T}]=
e^{-\Lambda T}\Expect \Bigl[  \frac{v(X_T)}{v(X_0)}  \exp\Bigl(\int_0^T  \util(X_t) \, dt\Bigr) \Bigr]
\]
The identity $\Lambda=\Lambda(\util)$ follows:  Since $v$ is strictly positive and finite-valued,  it follows that $\cp$ is infinite-horizon optimal:
\[
\lim_{T\to\infty} \frac{1}{T} \welf_T(\cp)  = \Lambda(\util) = 
\lim_{T\to\infty} \frac{1}{T} \Lambda^*_T  =
\welf_\infty^* 
\]
This establishes (ii).

Given the formula for $\cL_T$, the total welfare at time $T$ using $\cp$ is thus,
\[
 \begin{aligned}
\welf_T(\cp)  
&=     \Expect_{\cp}\Bigl[\int_0^T \util(X_t)\, dt\Bigr] -D(\cp_T \| p^0_T)
\\
  &=  - \Expect\Bigl[\log\Bigl( \frac{v(X_T)}{v(X_0)} \Bigr)  \Bigr]  + \Lambda T
\end{aligned}
\]
which establishes (iii).
\qed

\section{Mean field model and its linear approximation}
\label{s:meanfield}

Up to now we have only two generators:  The nominal generator $\generate$, and its transformation \eqref{e:cgenerate} obtained as the solution to an optimal control problem.
We next construct a parameterized family of generators denoted $\{\generate_\zeta :\zeta\in\Re\}$.
For each $\zeta$, this is obtained as   the infinite-horizon optimal control solution of the previous section, with the
finite-horizon welfare defined in \eqref{e:Twelfare}. 
If $\zeta=0$ then $ p^*=p^0$.  
For arbitrary $\zeta$,  the generator $\generate_\zeta$ that solves the infinite-horizon optimal control problem  is of the form \eqref{e:cgenerate}, in which $v=v_\zeta$ is a solution to the eigenvector problem,  
\begin{equation}
[\generate + \zeta I_{\util}] v_\zeta = \Lambda_\zeta v_\zeta
\label{e:evectorz}
\end{equation}
where  
\[\Lambda_\eta=
\Lambda(\zeta\util)
\eqdef
\lim_{T\to\infty} \frac{1}{T} \log \Expect\Bigl[ \exp\Bigl(  \zeta\int_0^T  \util(X_t) \, dt\Bigr) \Bigr] 
\]

The solutions to \eqref{e:evectorz} are used to define the continuous family of generators, $\generate_\zeta =   \ind_{v_\zeta}^{-1} [\generate +  \zeta I_{\util} - \Lambda_\zeta I]\ind_{v_\zeta}
$, or   componenet-wise, $
\generate_\zeta(x^i,x^j) =$
\begin{equation}   \frac{v_\zeta(x^j)}{v_\zeta(x^i)} \bigl[\generate(x^i,x^j) +  (\zeta\util (x^i) - \Lambda_\zeta )\ind\{x^j=x^i\} \bigr]
\label{e:cgenerate-zeta}
\end{equation}

The mean field model is the nonlinear state space model   defined by \eqref{e:muState},
which in this finite state space setting becomes,
\begin{equation}
\ddt \mu_t\, (x) = \sum_{x^i\in\state} \mu_t(x^i) \generate_{\zeta_t}(x^i,x)
\label{e:muStateFinite}
\end{equation}
Recall that $\pi$ denotes the unique   invariant probability measure for the nominal model.
The nominal model is called reversible if   the detailed-balance equations hold:
\[
\pi(x^i) \generate_0(x^i,x^j) = \pi(x^j) \generate_0(x^j,x^i) ,\qquad x^i,x^j\in\state 
\]

\begin{theorem}
\label{t:reversiblePassive}
Suppose that the nominal model is reversible.
Then its linearization \eqref{e:LSSmfg} satisfies,
\begin{equation}
\Real G(j\omega) =  \text{PSD}_Y(\omega)  ,\qquad \omega\in\Re\,,
\label{e:sweet}
\end{equation}
where 
\begin{equation}
\text{
$G(s) = C[Is-A]^{-1}B$ for $s\in\Co$.}
\label{e:G}
\end{equation}
\end{theorem}

The proof of the proposition involves a sequence of steps. 
The first steps are contained in \Section{s:resolve}:  The power spectral density for $\bfmY$ 
can be expressed in terms of the family of \textit{resolvent matrices} for the Markov process, and these can be interpreted as a component of the transfer function $G$.   The formula \eqref{e:sweet} is based on
these results, and a closer look at the linearization contained in \Section{s:lin}, which closes with a proof of  \Thm{t:reversiblePassive}.


Before proceeding with these technical arguments, we give an example to show that
the positive real condition
may not hold if the nominal model is not reversible.   
Moreover, this example shows that without reversibility, the linearization may not be minimum phase.

Consider the Markov chain with eight states, whose transition diagram and generator are shown in \Fig{fig:AB}.
This Markov process cannot be reversible because some transitions are uni-directional.  For example, an immediate transition from $3$ to $1$ is possible, but not from $1$ to $3$.  

In the notation of this paper we have $d=8$, and we take $x^i=i$ for $1\le i\le 8$.  
The utility function $\util\colon\state\to\Re$ is taken to be $\util(x)=x$.

\begin{figure}[h]
\Ebox{.65}{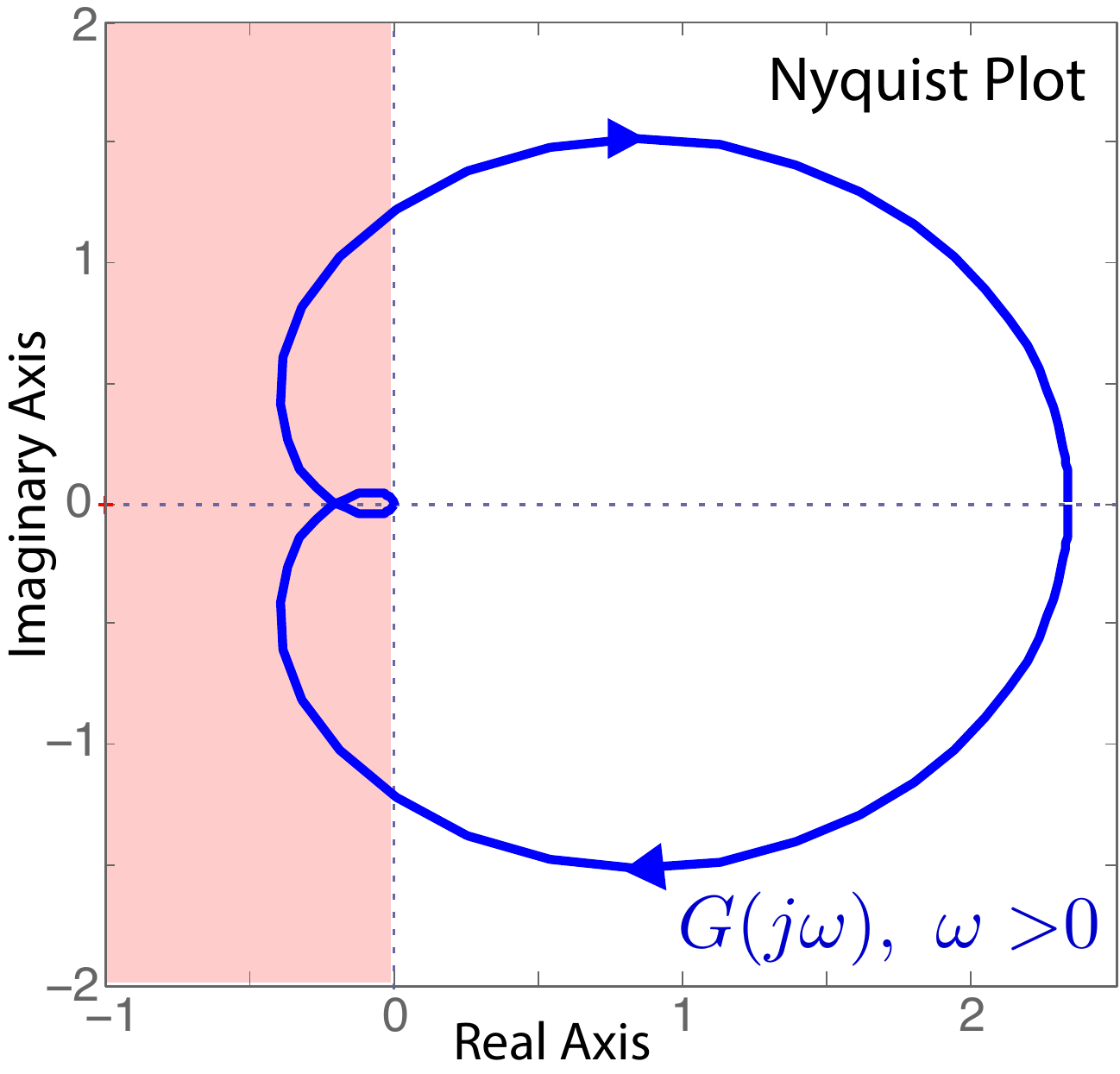}
\vspace{-.25cm}
\caption{Nyquist plot  for linearization} 
\label{fig:PoleZeroNyquist} 
\vspace{-.1cm}
\end{figure}

Shown in \Fig{fig:PoleZeroNyquist} 
 is a Nyquist plot 
 for the transfer function $G(s)= C[Is-A]^{-1} B$ with 
$a=c=10$ and $b=1$.
The Nyquist plot shows that the system is not positive real.  A pole-zero plot (not shown) reveals that the system is not minimum phase: there is a zero in the right half plane, at approximately $s=+9$.  

This example demonstrates that  the positive real condition obtained in
\Thm{t:reversiblePassive}
requires assumptions on the nominal model. Reversibility is used to obtain the identity \eqref{e:sweet} that implies the positive real condition.  We know of no  alternate assumptions that imply the positive-real condition.
 

\subsection{Resolvents and transfer functions}
\label{s:resolve}

The family of resolvent matrices are defined for   $s\in\Co$ by the integral,
\[
R_s = \int_0^\infty e^{-s t} P^t \, dt
\]
This is well defined whenever $\Real(s) < 0$.   It can be shown using \eqref{e:gendef} (or the representation $P^t=e^{t\generate}$)
that the resolvent equation holds $\generate R_s = s R_s -I$;  equivalently,
\begin{equation}
R_s =[s I - \generate]^{-1} 
\label{e:RsGen}
\end{equation}
We already see that this forms a component of $G(s)$ in \eqref{e:G} (recall that $A=\generate^\transpose$).   Consequently,    for each $s$ satisfying $\Real(s) < 0$,
\begin{equation}
G(s) = C[Is-A]^{-1}B = B^\transpose [Is-A^\transpose]^{-1}C^\transpose = B^\transpose R_s C^\transpose 
\label{e:GR}
\end{equation}
Based on this identity, the following result shows that the frequency response   is similar to a cross-power spectral density:


\begin{proposition}
\label{t:Ginterpret}
The frequency response for the transfer function \eqref{e:G} with $A=\generate^\transpose$ can be expressed,
for $\omega\in\Re$, by
\[
G(j\omega) 
=
\int_0^\infty e^{-j\omega t} \Expect_\pi[f(X_0) g(X_t)]\, dt  
\]
where $f(x^i) = B_i/\pi(x^i)$  and $g(x^i)=C_i=\tilutil(x^i)$,  $x^i\in\state$.
\end{proposition}

\proof
The proof begins with the representation \eqref{e:GR}, which holds by definition whenever $\Real(s) < 0$.  From the definition of the resolvent matrix,   \eqref{e:GR} gives,
\[
\begin{aligned}
G(s) & =  \int_0^\infty e^{-s t} \Bigl[\sum_{i,j} P^t(x^i,x^j)B_i C_j  \Bigr]\, dt   \\
& =
\int_0^\infty e^{-s t} \Expect_\pi[f(X_0) g(X_t)]\, dt  
\end{aligned}
\]
where the final equality follows from the definition of $f$ and~$g$.

To complete the proof we must extend  \eqref{e:GR} to $s=j\omega$, for which $\Real(s) = 0$.   For this we note that $g(x^i) = C_i =\tilutil(x^i)$,   so that 
$
\lim_{t\to\infty}\Expect_\pi[f(X_0) g(X_t)] = 0,
$
where the convergence rate is exponential. \notes{I can't find a reference!}
\qed

\subsection{Linearization}
\label{s:lin}

To apply Prop. \ref{t:Ginterpret} we require a representation of the matrix $B$ defined in \eqref{e:B}. 
For this we normalize the eigenvector so that $v_\zeta(x^1)=1$ for all $\zeta$;  this is without loss of generality since the components   $\generate_\zeta(x^i,x^j)  $ of the generator  \eqref{e:cgenerate-zeta} are defined in terms of the ratio $v_\zeta(x^j)/v_\zeta(x^i)$.

Let $h_0$ denote the solution to Poisson's equation,
\begin{equation}
\generate_0 h_0 = -\tilutil
\label{e:Fish}
\end{equation}
with boundary condition $h_0(x^1)=0$.  For a finite-state space Markov process, one solution to Poisson's equation is given by  
\begin{equation}
h(x)  = R_0 \tilutil \, (x) = \int_0^\infty  \Expect\bigl[ \tilutil(X_t) \mid X_0 =x\bigr]\, dt ,\quad x\in\state
\label{e:Fishpi0}
\end{equation}
and then we take $h_0(x) = h(x) - h(x^1)$,  $x\in\state$.

Let $ \generate^\adjsym $ denote the generator for the time-reversed process,
\[
\generate^\adjsym (x^i,x^j) =  \pi(x^j)  \generate (x^j,x^i) \frac{1}{\pi(x^i)},\qquad x^i,x^j\in\state
\]

\begin{proposition}
\label{t:B}
The entries of $B$ are given by,
\[
B_i =  
-\pi(x^i)
 [   \generate^\adjsym h_0\, (x^i)   + \generate h_0\, (x^i)   ]
\]
If the process is reversible, then $B_i = 2 \pi(x^i) C_i = 2 \pi(x^i)  \tilutil(x^i) $.
\qed
\end{proposition}

To prove the proposition we first need the following formulae for the derivatives of $\Lambda_\zeta$ and $v_\zeta$.  We omit the proof, which is similar to the discrete-time case \cite{konmey03a,konmey05a}. 
\notes{more precision would be nice}

\medskip

\begin{lemma}
\label{t:hder}
The log-moment generating function has derivative at the origin given by,
\[
\ddzeta \Lambda_\zeta \Big|_{\zeta=0} = \ynom = \sum_i \pi(x^i) \util(x^i)
\]
The derivative of the eigenvector is the solution to Poisson's equation,
\[
\ddzeta v_\zeta(x^i) \Big|_{\zeta=0}  = h_0(x^i),\qquad x^i\in\state.
\]
\qed
\end{lemma}

\paragraph{Proof of \Prop{t:B}}
Applying the lemma to
\eqref{e:cgenerate-zeta}
gives, for all $x^i,x^j\in\state$,
\[
\begin{aligned}
\generate_\zeta (x^i,x^j) 
&=  [1- \zeta h_0(x^i)] \generate (x^i,x^j) [1+ \zeta h_0(x^j)] \\
&  \quad +  [\zeta \util(x^i)   - \Lambda_\zeta ]\ind\{x^i=x^j\} +o(\zeta)
  \\
  &=  [1- \zeta h_0(x^i)] \generate (x^i,x^j) [1+ \zeta h_0(x^j)] \\
  & \quad +  \zeta [\util(x^i) - \ynom] \ind\{x^i=x^j\}     +o(\zeta)
\end{aligned}
\]
From the definition $\tilutil=\util-\ynom$  (see  \eqref{e:Cdefn}), we conclude that the derivative is given by,
\[
\begin{aligned}
\ddzeta \generate_\zeta (x^i,x^j)  \Big|_{\zeta=0} & =  -   h_0(x^i)  \generate (x^i,x^j) +   \generate (x^i,x^j) h_0(x^j) \\
& \; \quad  +    \tilutil(x^i) \ind\{x^i=x^j\}    
\end{aligned}
\]
The entries of the matrix $B$ are thus given by,
\[
\begin{aligned}
B_j &= \sum_{x^i} \pi(x^i) \Bigl(-   h_0(x^i)  \generate (x^i,x^j) +   \generate (x^i,x^j) h_0(x^j) \\
& \; \quad +    \tilutil(x^i) \ind\{x^i=x^j\}    
\Bigr)
\\
  &=   - \sum_{x^i}    h_0(x^i) \pi(x^i)  \generate (x^i,x^j)   + \pi(x^j) \tilutil(x^j) 
\end{aligned}
\]
where in the second identity we used the invariance equation, $\sum_{x^i} \pi(x^i)   \generate (x^i,x^j) =0$.  The second identity is equivalent to the desired representation.
\qed

 \medskip

\paragraph{Proof of \Thm{t:reversiblePassive}}
 \Prop{t:B} tells us that under reversibility we have  $B_i = 2 \pi(x^i)  \tilutil(x^i) $,
 and hence in the notation of \Prop{t:Ginterpret},
 \[
 f(x^i) = B_i/\pi(x^i) = 2 \tilutil(x^i),\qquad g(x^i)=C_i=\tilutil(x^i)
 \]
 \Prop{t:Ginterpret}  and
\Prop{t:B} then give,
\[
 \begin{aligned}
G(j\omega) 
&=
\int_0^\infty e^{-j\omega t} \Expect_\pi[f(X_0) g(X_t)]\, dt  
\\
&=
2\int_0^\infty e^{-j\omega t} \Expect_\pi[\tilutil(X_0) \tilutil(X_t)]\, dt  
\end{aligned}
\]
\Thm{t:reversiblePassive}
thus follows:
\[
 \begin{aligned}
 \Real
G(j\omega)  
&=
2\Real\int_0^\infty e^{-j\omega t} \Expect_\pi[\tilutil(X_0) \tilutil(X_t)]\, dt  
\\
&=
 \int_{-\infty}^\infty e^{-j\omega t} \Expect_\pi[\tilutil(X_0) \tilutil(X_t)]\, dt  
\end{aligned}
\]
\qed


\section{Conclusions}
\label{s:conc}

This paper gives a general condition under which the linearization of a mean field model is positive-real. 

The linearization around $\zeta=0$ is a natural choice, but the main result of the paper can be extended to any constant value:  If $\generate $ is reversible, then so is $\generate_\zeta$ for each fixed $\zeta\in\Re$.   Based on this observation, it is possible to show that the linearization about any fixed value of $\zeta $ is positive real under the assumptions of \Thm{t:reversiblePassive}.
This suggests an open question:  Is the nonlinear model with state equation \eqref{e:muState}
 passive?  
Passivity would be a valuable property for the purposes of control.
 
There are many open questions in the context of design.  Can we obtain more general sufficient conditions for the positive real condition, the weaker minimum phase condition, or the stronger  passivity condition for the nonlinear model?

To relax the assumptions of \Thm{t:reversiblePassive}, it is likely that we will require application of the Kalman-Yakubovich-Popov Lemma, which provides an algebraic characterization of the passive real condition \cite{ran96}.
 
We are currently considering these theoretical questions, and applications to problems in decentralized control, especially in power systems settings.

\notes{say somewhere:  The minimum phase condition observed in the numerical examples in \cite{meybarbusyueehr14} cannot be explained by any of the theory in this paper
since a Markov model with state transition diagram shown in \Fig{fig:pppDynamics} 
 cannot be reversible.}

%
 
 \newpage
 
 \def\cprime{$'$}\def\cprime{$'$}

\end{document}